\documentstyle[prl,aps]{revtex}
\draft
\input epsf
\input{epsf.sty}
\begin{document}

\twocolumn[\hsize\textwidth\columnwidth\hsize\csname
@twocolumnfalse\endcsname

\title{
Intrinsic Percolative Superconductivity in Heavily Overdoped High Temperature Superconductors}

\author{
H. H. Wen\cite{responce}, S. L. Li, Z. W. Zhao, Z. Y. Liu, H. P. Yang and D. N. Zheng
}

\address{
National Laboratory for Superconductivity,
Institute of Physics and Center for Condensed Matter Physics,
Chinese Academy of Sciences, P.O. Box 603, Beijing 100080, China}

\maketitle

\begin{abstract}
Magnetic measurements on heavily overdoped $La_{2-x}Sr_xCuO_4$, $Tl_2Ba_2CuO_6$, $Bi_2Sr_2CuO_6$ and $Bi_2Sr_2CaCu_2O_8$ single crystals reveal a new type magnetization hysteresis loops characterized by the vanishing of usual central peak near zero field. Since this effect has been observed in various systems with very different structural details, it reflects probably a generic behavior for all high temperature superconductors. This easy penetration of magnetic flux can be understood in the picture of percolative superconductivity due to the inhomogeneous electronic state in heavily overdoped regime.      
\end{abstract}


]
The mechanism of high temperature superconductors ( HTS ) remains puzzling and one of the most attractive topics in condensed matter physics. An essential focus lies on the understanding to the parabolically shaped superconducting area in the generic electronic phase diagram containing several phases depending on the hole doping level: underdoped, optimally doped and overdoped. The contrasting properties in the normal state between an underdoped and an overdoped sample tempt to ascribe the superconductiviy to different condensation processes. For example, one picture suggests that the superconducting transition is a Bose-Einstein condensation in the underdoped region and the BCS type condensation in the overdoped region \cite{uemura1,deutscher}. The crossover from the non-Fermi liquid in the underdoped region to the approximate Fermi liquid behavior in the overdoped region may manifest that most of the doped holes join the conduction in the normal state. Recent data from the measurement on the penetration depth $\lambda$ \cite{bernhard}, show that, however, the superfluid density $\rho_s$ drops with the doping level p when $ p > 0.19 $. This leads to a difficult point that in the heavily overdoped regime, the more charge carriers are doped, the less superfluid density $\rho_s$ will be. Therefore, the doped holes in the overdoped region seem to be separated into two parts, only part of them will condense into a lower energy state leading to the superconductivity. 

In our previous papers \cite{wen1}, we have presented a preliminary evidence to show that the electronic state in the overdoped region may be inhomogeneous \cite{gim}. According to this picture, superconducting condensation will occur first in some local region ( with less holes ) to form some tiny superconducting islands. The extra holes on these superconducting islands will be expelled to the surrounding area to form a hole-rich non-superconducting metallic sea and the bulk superconductivity will be established by the Josephson coupling or proximity effect between these islands. This picture is different from the Swiss-Cheese model proposed early by Uemura \cite{uemura2} which involves only a microscopic inhomogeneity. Therefore it warrants always further investigations until it is finalized with solid experimental evidence. If this picture is correct, one should be able observe the percolative superconductivity in the heavily overdoped region. In this Letter, for the first time, we present a new type magnetization hysteresis loop ( MHL ) characterized by the vanishing of the central penetration peak. This effect can be explained quite well by the picture based on the percolative superconductivity.

Single crystals measured for this work were prepared by the traveling solvent floating-zone technique (  LSCO, Bi-2201 ) and the self-flux method ( Tl-2201 and Bi-2212 ). Several single crystals have been investigated for this study. Our major conclusion here has been re-checked by using many crystals of LSCO, Tl-2201, Bi-2201 and Bi-2212. The data from Bi-2201 and Bi-2212 will not be shown here due to the length limit of this paper. All these single crystals have rather narrow superconducting transitions as observed from the temperature dependent magnetization measurement. The X-ray diffraction pattern ( XRD ) taken from these single crystals show very good crystallinity without any trace of a second phase. All these guarantee the generality of the major conclusion drawn in this paper. A Quantum Design superconducting quantum interference device ( SQUID, MPMS 5.5 T ) and an Oxford vibrating sample magnetometer ( VSM 3001, 8 T ) were used to measure the MHL.

For an uniform suerconductor in the superconducting state, the condensate of the superconducting electrons will expel the external field. When the external field $ H $ is higher than the lower critical field $ H_{c1} $, many quantized magnetic vortices will be formed and penetrate into the interior of the sample. The spatial distribution of the density of these vortices or called as flux profile is illustrated in the inset of Fig.1. The curve of M vs. H will deviate from the linear relation $M = -H/4\pi$ ( Meissner state) at $ H_{c1} $ for a perfect cylinder. While the magnetization M will continue to grow until the flux front meets at the center of the sample ( $ B_e = B_p$ ). By further increasing the external field more and more magnetic 
\begin{figure}[h]
    \vspace{10pt}	
    \centerline{\epsfxsize 8cm \epsfbox{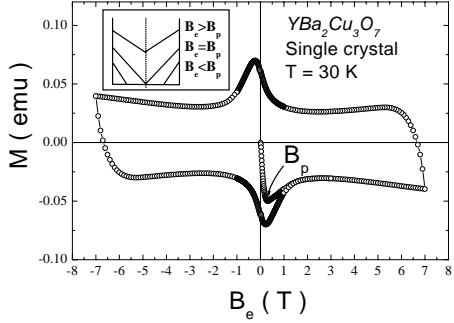}}
    \vspace{10pt}
\caption{A typical MHL measured for a $YBa_2Cu_3O_{7-\delta}$ single crystal at T = 30 K. A clear penetration peak appears near zero field. This central penetration peak remains until $ T \geq T_c$. The inset shows the magnetic flux profile in a superconducting cylinder at different fields. When $ B_e = B_p $, the magnetization reaches the maximum.
}
\label{fig:Fig1}
\end{figure}
\noindent vortices will creep into the sample, the magnetization will start to drop. Therefore for any uniform superconductor, a penetration peak near zero field will appear due to the non-easy penetration of magnetic flux. As an example, in Fig.1 we show one MHL for an $ YBa_2Cu_3O_{7-\delta} $ single crystal near optimal doping. A clear penetration peak can be observed here. When the field drops down from a high value ( from both positive and negative side ), a huge magnetization peak will appear near zero field due to establishing a large superconducting current. This huge central peak envelops the penetration peak. One can imagine that the penetration peak and the central peak may disappear for a percolative superconducting system due to the difficulty of establishing a high current density near zero field.   

As the magnetic flux fully penetrate into the sample, some vortex phases can be formed in some specific temperature and field region. For example, a picture based on the crossover from Bragg glass in low field region to the vortex glass in high field region \cite{giamarchi} was proposed to interpret the second peak effect \cite{bi2212} ( narrow and sharp but almost temperature independent ) in Bi-2212 single crystals as shown in Fig.2. For $ YBa_2Cu_3O_{7-\delta} $ single crystals, a broad and temperature dependent second peak \cite{ybco} has been observed which has received two major explanations\cite{daeumling,krusin-elbaum}. However, \textit { for any underdoped or optimally doped high temperature superconductors no matter how low the doping level or superfluid density is, and no matter what kind of the shape of the sample, to our knowledge, on the MHL curve, there exists a central penetration peak due to the non-easy penetration of the magnetic flux}.

For a heavily overdoped HTS, the situation could be different according to the picture mentioned above. Thus we have intentionally investigated the magnetic flux 
\begin{figure}[h]
    \vspace{10pt}	
    \centerline{\epsfxsize 8cm \epsfbox{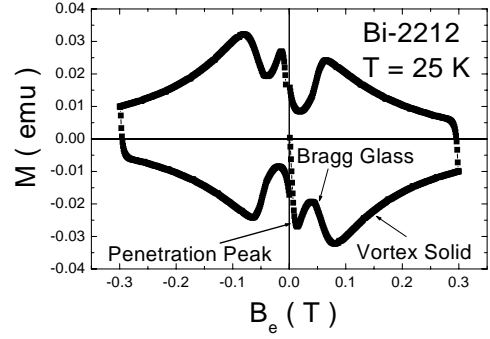}}
    \vspace{10pt}
\caption{A typical MHL measured for an optimally doped Bi-2212 single crystal ( $T_c$ = 93 K ) at T = 25 K. A clear penetration peak can be observed near zero field. This central penetration peak remains until $ T \geq T_c$. At a higher field, a second peak appears which is explained as the crossover from the low field Bragg glass to the high field vortex glass.
}
\label{fig:Fig2}
\end{figure}
\noindent penetration process for such samples. In Fig.3(a), a typical MHL measured at 9 K for an overdoped LSCO ( x = 0.24 ) is shown. Clearly the central peak becomes too small to be observable. From an enlarged view shown in the inset to Fig.3(a) one can still see a very tiny central peak. If comparing this MHL to that measured for $ YBa_2Cu_3O_{7-\delta} $, Bi-2212 and underdoped LSCO ( x = 0.092, $T_c$ = 26 K, shown in Fig.3(b)), one can clearly see that the central peak becomes very small or disappears for the overdoped LSCO ( x = 0.24 ) sample. 

For our present overdoped LSCO single crystal ( x = 0.24, $ T_c$ = 25 K ), it is known that $t_c = T_c/T_c^{opt} $ = 25 K / 38 K = 0.66, thus the sample is close to the heavily overdoped region. In Fig.4(a) and (b), we present the MHLs for a Tl-2201 single crystal ( $ T_c $ = 35 K ) with $ t_c = T_c / T_c^{opt}$ = 35 K / 95 K = 0.37, thus it is in the heavily overdoped regime. In the low temperature region, a penetration peak is observed at a rather low field. This penetration peak at a low field is resulted from the relatively strong coupling between the tiny superconducting islands in low temperature region. When the temperature is increased, this first peak disappears quickly ( as shown in Fig.4 (b)) and a second peak at a higher field develops. This again shows the easy penetration of magnetic flux in the heavily overdoped region.  

Next we present further evidence by showing the dynamical flux penetration process. According to our picture, the bulk superconducting state is established by the Josephson coupling between the tiny superconducting islands. In short time scale the high slope of flux gradient $dB(x)/dx$ corresponding to a large supercurrent can be maintained since the state relaxes only a little from the initial state ( t = 0, the moment the field sweeping is just stopped  ). A long time relaxation process will drive the system more close to the static state so that the central  
\begin{figure}[h]
    \vspace{10pt}	
    \centerline{\epsfxsize 8cm \epsfbox{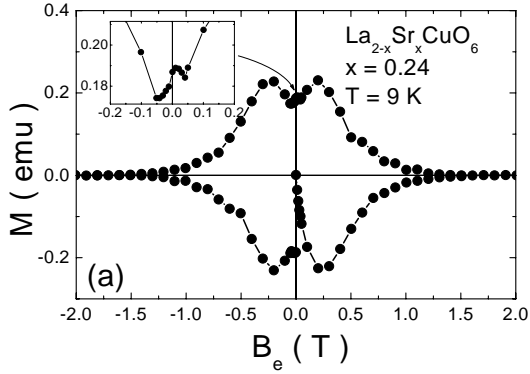}}
    \vspace{10pt}
    \vspace{10pt}	
    \centerline{\epsfxsize 8cm \epsfbox{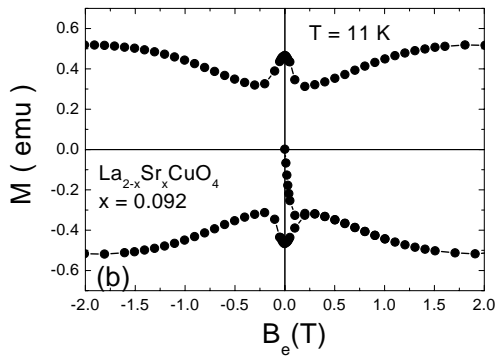}}
    \vspace{10pt}
\caption{MHLs measured for (a) an overdoped ( x =0.24, $T_c$ = 25 K ) and (b) an underdoped ( x = 0.096, $ T_c$ = 26 K ) LSCO single crystals. For the underdoped sample, a clear central penetration peak is observed. While for the overdoped sample, the central peak is too small to be observable showing an easy penetration process.
}
\label{fig:Fig3}
\end{figure}
\noindent penetration peak will become more and more invisible. In order to check this idea, we have measured the time dependent magnetization from 0.01 s to 3600 s. From these data we construct a portion of the MHL measured at different time scales. As shown in Fig.5, at a short time the magnetization near zero field shows a plateau with a trace of a small central peak. However, it relaxes quickly versus time leading to the vanishing of the central peak. One can expect that at a shorter time scale, the central peak will become more clear.

There are several possibilities for explaining the vanishing of central peak on MHL. The first naive argument would be that this effect is due to the chemical inhomogeneity. However, this argument stands weakly against the very pure and neat structural data from XRD, TEM and other analyses. It is also inconsistent with the common observation of this effect in various system with very different structural details. For example, in LSCO single crystals there is a structural phase transition from orthorhombic to tetragonal at a certain temperature.  
\begin{figure}[h]
    \vspace{10pt}	
    \centerline{\epsfxsize 8cm \epsfbox{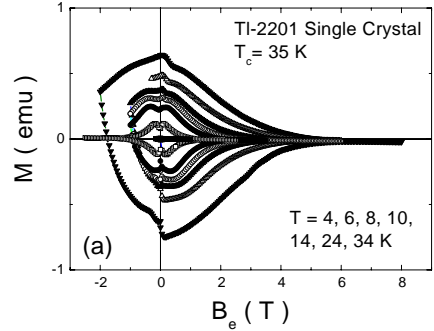}}
    \vspace{10pt}
    \vspace{10pt}	
    \centerline{\epsfxsize 8cm \epsfbox{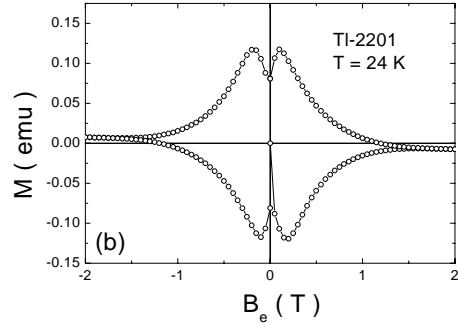}}
    \vspace{10pt}
\caption{MHLs measured for a heavily overdoped Tl-2201 single crystal at (a) temperatures of 4, 6, 8, 10, 14, 24, 34 K ( from outer to inner ) and (b) at 24 K. In low temperature region, a penetration peak appears at a rather low field. This peak at a low field is probably resulted from the relatively strong coupling between the tiny superconducting islands at low T. When T is increased, this first penetration peak disappears.}
\label{fig:Fig4}
\end{figure}
\noindent While in Tl-2201, Bi-2201 and Bi-2212 systems no such transition is expected. In addition, the doping to the overdoped region in LSCO and Bi-2201 system is by cation substitution, while in Tl-2201 and Bi-2212 is through post-annealing in oxygen leading to rather different structural deformations. When a optimally doped Bi-2212 sample is post-annealed in high pressure $O_2$ ( 100 atm. 500$^\circ C$ ), $T_c$ drops from 95 K to 72 K and the central peak becomes much smaller. Similar data were published recently \cite{bi2212over} although no clear reasons have been given. It would be very interesting to see whether this central peak will vanish completely when further doping Bi-2212 to heavily overdoped regime. Since this effect has been found in four systems: LSCO, Tl-2201, Bi-2201 and Bi-2212, we anticipate that whenever a HTS is driven to the heavily overdoped region one should be possible to see this effect. Work on the Ca doped $YBa_2Cu_3O_7$ (  another typical overdoped sample ) is under way.

\begin{figure}[h]
    \vspace{10pt}	
    \centerline{\epsfxsize 8cm \epsfbox{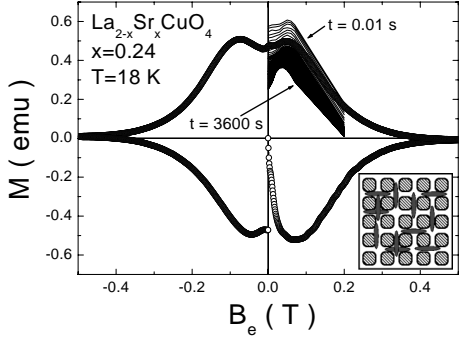}}
    \vspace{10pt}
\caption{Time dependent MHLs for the overdoped LSCO single crystal ( x = 0.24, $ T_c = 25 K $ ). It is clear that the magnetization at zero field relaxes much faster than that at a high field leading to the vanishing of the central penetration peak. The inset gives a schematic show for the percolating superconductivity in the heavily overdoped regime. The round square and the ellipse stand for the tiny superconducting islands and the Josephson vortices respectively. }
\label{fig:Fig5}
\end{figure}
\noindent 

The second argument would be that this effect is induced by very low superfluid density $\rho_s$ although the superconducting state is uniform \cite{stripe}. This argument gets, however, no support from the data measured for \textit{any} underdoped HTS sample which has also a very low $\rho_s$. As mentioned already, this easy penetration effect has never been observed in underdoped HTS samples. Therefore even for a superconductor with very low superfluid density, one should be able to observe a penetration peak when the superconducting state is uniform.

This effect can get an explanation based on the electronic phase separation in heavily overdoped HTS sample. The resulting phase consists many tiny superconducting islands surrounded by the hole-rich metallic but non-superconducting regions. The bulk superconducting state is established by the Josephson coupling or proximity effect between these islands. It has been found that the superconducting criticality for the underdoped and the overdoped sample is the same \cite{wen2} reflecting probably the same condensation processes, e.g., both are BEC in the two doping limits. A schematic show for this picture is given in the inset to Fig.5. In the low field region, the magnetic flux ( probably the Josephson vortices ) will easily penetrate into the center of the sample through these weak coupling channels. Therefore the central peak will become too small to be observable. Since the decoupling transition coincides with the irreversibility line \cite{wen1,wen2}, we would not expect any irreversible flux motion within the superconducting islands. This may manifest that the islands are too small to allow the formation of vortex lattice. In this case, we can expect that the islands have a size between several times of coherence length and 1000 $ \AA $. When field is increased, one can observe an increase of the magnetization. This may be induced by the gradual formation of a vortex solid state. However, it is important to note that this vortex solid state, if exists, is formed by the Josephson vortices rather than the normal vortices. This deserves certainly a further check.

In conclusion, by measuring the MHLs for several heavily overdoped high temperature superconductors ( LSCO, Tl-2201, Bi-2201 and Bi-2212 ), for the first time we have observed a new type MHL characterized by the vanishing of central penetration peak. This effect has never been observed on underdoped or optimally doped samples and can be attributted to the percolative superconductivity due to inhomogeneous electronic state in heavily overdoped region.

\acknowledgements
This work is supported by the National Science Foundation of China (NSFC 19825111) and the Ministry of Science and Technology of China ( project: NKBRSF-G1999064602 ). We acknowledge helpful discussions with Prof. Duan Feng at Nanjing university, China, Dr. Thierry Giamarchi ( Paris ) and Dr. Xiao Hu ( Trsukuba, Japan ).

\end{document}